\documentclass[conference]{IEEEtran}
\usepackage[backend=biber,style=ieee,doi=false,url=false,
            citestyle=numeric-comp,maxbibnames=3]{biblatex}
\AtBeginBibliography{\footnotesize}
\usepackage{amsmath,amssymb,amsfonts}
\usepackage{algorithmic}
\usepackage{graphicx}
\usepackage{textcomp}
\usepackage{xcolor}
\usepackage{xspace}
\usepackage{csquotes}
\usepackage{comment}
\usepackage{setspace}
\usepackage{etoolbox}
\usepackage{nicefrac}
\usepackage[inline]{enumitem}
\usepackage[colorlinks=true,urlcolor=teal,
            linkcolor=teal,citecolor=teal]{hyperref}
\usepackage{booktabs}
\usepackage{mathtools}\mathtoolsset{showonlyrefs}
\usepackage{siunitx}
\usepackage{subcaption}
\usepackage{tikz}

\newcommand{\reprourl}{https://github.com/lfd/wihpqc2024-noisy-qaoa}
\newcommand{\zenodourl}{https://doi.org/10.5281/zenodo.12792769}
\newcommand{\repro}{\href{\reprourl}{reproduction package}}

\usepackage{dsfont}
\newcommand{\bra}[1]{\langle #1 |}
\newcommand{\ket}[1]{| #1 \rangle}

\newcommand{\imag}{\mathrm{i}}

\def\subsectionautorefname{section}

\newcommand{\secref}[1]{%
\begingroup%
\def\chapterautorefname{Chapter}%
\def\sectionautorefname{Section}%
\def\subsectionautorefname{Section}%
\autoref{#1}%
\endgroup%
}

\newcommand{\eg}{\emph{e.g.},\xspace}
\newcommand{\ie}{\emph{i.e.},\xspace}
\newcommand{\etal}{\emph{et al.}\xspace}

\newcommand{\cnot}{\ensuremath{\text{C\raisebox{0.08em}{--}}\!X}\xspace}

\newcommand{\rz}{R_Z}

\usepackage{censor}
\newboolean{anonymous}
\setboolean{anonymous}{false}
\ifbool{anonymous}{}{\renewcommand{\censor}[1]{#1}}
\ifbool{anonymous}{}{\renewcommand{\blackout}[1]{#1}}
\ifbool{anonymous}{\newcommand{\genemail}[2]{\href{mailto:xxx.xxx@xx.xx}{\blackout{#2}}}}{\newcommand{\genemail}[2]{\href{#1}{#2}}}

\newcommand{\tikzspace}{\vspace{-0.3cm}}

\AtEveryBibitem{%
  \clearfield{note}%
}
\AtEveryBibitem{\clearfield{month}}
\AtEveryBibitem{\clearfield{day}}
\AtEveryBibitem{\clearfield{date}}
\AtEveryBibitem{\clearfield{issn}}
\AtEveryBibitem{\clearfield{isbn}}
\AtEveryBibitem{\clearfield{address}}
\AtEveryBibitem{\clearfield{editor}}

\newbibmacro{string+doi}[1]{%
  \iffieldundef{doi}{#1}%
  {\href{http://dx.doi.org/\thefield{doi}}{#1}}}
\DeclareFieldFormat*{title}{\usebibmacro{string+doi}{\mkbibemph{#1}}}

\begin{document}
\title{Approximating under the Influence\\ of Quantum Noise and Compute Power}

\author{
    \IEEEauthorblockN{\blackout{Simon Thelen}}
    \IEEEauthorblockA{
        \blackout{\textit{Technical University of}} \\
        \blackout{\textit{Applied Science Regensburg}} \\
        \blackout{Regensburg, Germany} \\
        \genemail{mailto:simon.thelen@othr.de}{simon.thelen@othr.de}
    }
    \and
    \IEEEauthorblockN{\blackout{Hila Safi}}
    \IEEEauthorblockA{
        \blackout{\textit{Siemens AG, Technology}} \\
        \blackout{\textit{Technical University of}} \\
        \blackout{\textit{Applied Science Regensburg}} \\
        \blackout{Munich, Germany} \\
        \genemail{mailto:hila.safi@siemens.com}{hila.safi@siemens.com}
    }
    \and
    \IEEEauthorblockN{\blackout{Wolfgang Mauerer}}
    \IEEEauthorblockA{
        \blackout{\textit{Technical University of}} \\
        \blackout{\textit{Applied Science Regensburg}} \\
        \blackout{\textit{Siemens AG, Technology}} \\
        \blackout{Regensburg/Munich, Germany} \\
        \genemail{mailto:wolfgang.mauerer@othr.de}{wolfgang.mauerer@othr.de}
    }
}

\maketitle

\begin{abstract}
  The quantum approximate optimisation algorithm (QAOA) and its variants are at the core of many scenarios that aim at combining the power of quantum computers (QC) and classical high-performance computing (HPC) appliances for combinatorial optimisation. Several obstacles challenge concrete benefits now and in the foreseeable future: Imperfections quickly degrade algorithmic performance below practical utility; overheads arising from alternating between classical and quantum primitives can counter any advantage; and the choice of parameters or  algorithmic variant can substantially influence runtime and result quality. Selecting the appropriate combination is a non-trivial issue, as it not only depends on
  end-user requirements, but also on details of the hardware and software stack.
  Appropriate automation can alleviate the burden of choosing optimal combinations for end-users: They should not be required to understand technicalities like  detail differences between QAOA variants, required number of QAOA layers, or necessary measurement samples. Yet, they should receive the best possible satisfaction of their non-functional requirements, be it performance or other. 
  We determine factors that influence approximation quality and temporal behaviour of four QAOA variants using comprehensive density-matrix-based numerical simulations targeting three widely studied optimisation problems. Our simulations consider ideal quantum computation, and a continuum of scenarios troubled by realistic imperfections.
  
  Our quantitative results, accompanied by a comprehensive reproduction package, show strong performance differences between QAOA variants
  that can be pinpointed to narrow and specific effects.
  We identify influential co-variables and relevant non-functional quality goals that, as we argue, mark the relevant ingredients for designing appropriate software engineering abstraction mechanisms  and automated tool-chains for devising quantum solutions from higher-level problem specifications.
\end{abstract}


\section{Introduction}
Quantum computers have the potential to solve certain problems exponentially faster than classical computers.
However, the capabilities of current noisy intermediate-scale quantum~(NISQ) systems are limited by the influence of quantum noise that severely disturbs any quantum computation~\cite{nisq}, and by the small number of available qubits, which inhibits quantum error correction~\cite{error-correction} for now.
Given the quantum-classical structure of NISQ-era algorithms, it is plausible that rather than replacing existing systems, quantum computers will be integrated into HPC (or embedded or special-purpose) architectures to accelerate specific tasks such as optimisation or simulation.
This hybrid approach will allow HPC systems to leverage the strengths of both, quantum and classical computing, and will likely extend into the post-NISQ era.
Variational quantum algorithms such as the quantum approximate optimisation algorithm (QAOA) are at the forefront of current research~\cite{qaoa}.
QAOA uses a multi-layer, structured variational quantum circuit to find approximate solutions for a broad class of optimisation problems with many industrial applications.
The more layers are used, the better the algorithm can approximate the optimal solution under ideal conditions.
However, a deeper circuit on NISQ systems also increases the number of circuit parameters, execution time and, perhaps most importantly, susceptibility to noise.
To complicate matters further, HPC engineers can choose between different QAOA variants such as warm-starting QAOA (WSQAOA) \cite{wsqaoa} or recursive QAOA (RQAOA) \cite{rqaoa,rqaoa2}.
These address noise by putting a stronger focus on the classical part of the algorithm.

Depending on the specific use-case, different non-functional requirements such as required solution quality, execution time or noise resistance must be taken into account.
This results in a number of potential trade-offs that must be carefully considered to put QAOA to good use in HPC environments.
It is not  obvious a-priori which variant performs best for which task, given the constraints of NISQ hardware and the desired result quality.
Ideally, selecting a suitable QAOA variant should be done automatically by the compiler or runtime environment.
However, the effects of noise on QAOA are not fully understood, especially for non-standard variants of the algorithm. To the best of our knowledge, no comprehensive study compares different QAOA variants under noise.
Our work, which is accompanied by a \href{\reprourl}{code repository} and a \href{\zenodourl}{reproduction package} (links in PDF),
aims to address this knowledge gap by examining the effect of factors like QAOA variant, number of layers, and noise strength on the non-functional properties of programs in HPC systems augmented with NISQ hardware. We use density-matrix-based numerical simulations to analyse solution quality and execution times of four QAOA variants on three widely studied optimisation problems.
Our objective is to identify factors with the strongest influence on non-functional quality goals, to guide the design and development of appropriate abstraction mechanisms and automatic tools based on high-level problem descriptions.

\section{Context and Foundation}\label{sec:foundation}
\subsection{Subject Problems}
We study the performance of multiple QAOA variants using three NP-complete optimisation problems: Max-Cut (which seeks to  partition the vertices of an undirected graph into two disjoint sets $S$ and $T$ such that the number of edges between $S$ and $T$ is minimised), Partition (which seeks to distribute a set of numbers into disjoint subsets $S$ and $T$ such that the absolute difference between the sum of numbers in $S$ and $T$ is minimised) and Vertex Cover (which seeks to find the smallest vertex subset $C$ of an undirected graph such that for every edge $(u, v)$ in the graph, either $u \in C$ or $v \in C$).
The subject problems
\begin{enumerate*}[label=(\alph*)]
\item are well-understood, with many applications, 
\item have efficient encodings with only one qubit per vertex (Max-Cut, Vertex Cover) or number (Partition), and 
\item differ considerably in their hardness of approximation: A fully polynomial-time approximation scheme is known for Partition~\cite{subset-sum-fptas}, but the best-known approximation ratios are 0.878 for Max-Cut~\cite{gw-rounding} and 2 for Vertex Cover~\cite{vertex-cover-approximation}.
\end{enumerate*}

\subsection{QAOA and its variants}
\label{subsec:qaoa}
QAOA finds approximate solutions to combinatorial optimisation problems~\cite{qaoa}.
The basic algorithm seeks parameters $s$ to minimise $C(s) = -\sum_{i < j} J_{ij} s_i s_j - \sum_i h_i s_i$ of a quadratic binary unconstrained objective function (QUBO)
with $s_i = \pm 1$ and $J_{ij}, h_i \in \mathbb{R}$.
Many NP-complete optimisation problems, including our subject problems, have efficient QUBO encodings~\cite{ising}.
A QAOA circuit first prepares an initial state, typically  $\ket{+}^n$, and applies a series of unitaries: $e^{-\imag \beta_p H_M} e^{-\imag \gamma_p H_C} \dots e^{-\imag \beta_1 H_M} e^{-\imag \gamma_1 H_C}$, where \(p\) chooses the number of \emph{layers}.
$H_C$ is the \emph{problem Hamiltonian} (or separator, as it separates solution space from search space by a 
complex phase), with $H_C\ket{x} = C((-1)^{x_1}, \dots, (-1)^{x_n}) \ket{x}$.
$H_M = \sum_i X^{(i)}$ is called the \emph{mixer Hamiltonian}, which influences the quantum state to explore the search space.
Optimal parameters $\beta_i, \gamma_i$ ($1 \leq i \leq p$) are found through multiple circuit evaluations using a classical optimiser.
More layers improve, in principle, results at the expense of runtime, but also amplify NISQ noise, which decreases solution quality.

Warm-starting QAOA variants alter the initial state and/or mixer Hamiltonian 
using a known initial guess for the solution.
This can improve result quality, especially at low depths~\cite{wsqaoa}.
We consider two variants:
For WS-Init-QAOA, given an approximate solution $z^* \in \{-1, +1\}^n$, we use $\bigotimes_i RY(\theta_i) \ket 0$ with $\theta_i = 2 \arcsin\left(\sqrt{0.5 - 0.25 z^{*}_i}\right)$ as the initial state instead of $\ket{+}^n$ \cite{wsinitqaoa}.
For WSQAOA, in addition to the altered initial state, we also change the time-evolved mixer Hamiltonian to $\bigotimes_i RY(-\theta_i) RZ(-2\beta) RY(\theta_i)$ \cite{wsqaoa}.

Recursive QAOA (RQAOA)~\cite{rqaoa,rqaoa2}
samples the circuit with the final parameters $\vec{\beta}$ and $\vec{\gamma}$ after QAOA 
execution. Then, the individual terms of the Hamiltonian $T = \{s_i : h_i \neq 0\} \cup \{s_i s_j : J_{ij} \neq 0\}$ are considered:
Term $t \in T$ with the largest absolute expected value $|E[t]|$ according to sampling results
is selected, and constraint $t = \operatorname{sign}(E[t])$ is inserted into the Hamiltonian by substituting $s_i = \pm 1$ or $s_i = \pm s_j$, which eliminates one variable.
This process of using QAOA to find the \enquote{most conclusive} term is iterated until the problem becomes trivial to solve.
Finally, a solution that satisfies all encountered constraints is selected.

\section{Related Work}\label{sec:related-work}
QAOA~\cite{qaoa} is among the most prevalent algorithms considered for NISQ devices. A wide range of efforts focuses on optimising QAOA for available quantum hardware, particularly to mitigate the effect of high error rates. Modifications to both,  structure of the quantum circuit (\eg Refs.~\cite{Zhang:2017,Wang:2020,Baertschi:2020,rqaoa,Zhu:2022}) and classical optimisation procedure (\eg Refs.~\cite{wsqaoa,wsinitqaoa,Tate:2023,Vijendran:2024,Sud:2024,montanezbarrera:2024,Streif:2020}) have been proposed. We cannot consider the abundance of choices, but focus on typical
variants, as described above: WSQAOA~\cite{wsqaoa}, WS-Init-QAOA~\cite{wsinitqaoa} and RQAOA~\cite{rqaoa}.

As for quantum noise, Georgopoulos~\etal~\cite{noise-model} present an approach to simulate effects of three error types using quantum channels, and align the model with experimental observations. Greiwe~\etal~\cite{felix} investigate the effects of imperfections on quantum algorithms. 
Xue~\etal~\cite{noisy-qaoa} confirm the effectiveness of hybrid algorithms on NISQ devices by studying effects of quantum noise on standard QAOA. Marshall~\etal~\cite{noisy-qaoa2} provide an approximate model for fidelity and expected cost given noise rate, system size, and circuit depth. 
Integrating QC in HPC environments poses  challenges, and comprehensive software stacks to address these are being designed and implemented~\cite{Humble:2021,Alexeev:2024,Karalekas:2020,Bandic:2022,Farooqi:2023,Elsharkawy:2023,Campbell:2023}; alas, we cannot review all in detail.
Bandic~\etal~\cite{Bandic:2022} survey QC full-stacks, highlighting the need for tight co-design and vertical integration between software and hardware. Auto-tuning 
in HPC environments was studied by Hoefler~\etal~\cite{Hoefler:2015}.
Wintersperger~\etal~\cite{Wintersperger:2022} and Safi~\etal~\cite{safi:23:codesign} study the influence of parameters like communication latencies and adapted topologies in HPC/QC systems. 
Elsharkawy~\etal~\cite{Elsharkawy:2023} assess the suitability of quantum programming tools for integration with classical HPC frameworks.
Close integration of classical and quantum aspects is a paramount desire in most studies.

\section{Experimental Setup}
\label{sec:experimental-setup}
We compare the ideal and noisy performance of different QAOA variants using density matrix-based, numerical simulations. We use the \href{https://atos.net/en/solutions/quantum-learning-machine}{Eviden Qaptiva 800} quantum simulation platform and its proprietary software library \emph{QLM} that includes a high-performance noisy circuit simulator.
Different numbers of layers ($p \in \{1,2,3,4\}$) are investigated for each variant. 
We will analyse performance on random instances of the three problems \emph{Max-Cut}, \emph{Partition} and \emph{Vertex Cover}:
For Max-Cut and Vertex Cover, 600 random graphs are considered, 100 for each $n \in \{5, 6, 7, 8, 9, 10\}$.
These graphs are created by inserting an edge between every pair of vertices with probability $p = 0.5$.
For Partition, 100 sets of numbers are considered for every size $n \in \{5,6,7,8,9,10\}$.
Each number is drawn uniformly at random from the interval $[0, 1]$.
To obtain more accurate data, the algorithms are run multiple times on each instance, averaging the results.
For all variants, SciPy's COBYLA optimiser with a tolerance of \SI{1}{\%} and 150 maximum iterations is used as the classical parameter optimiser~\cite{scipy,cobyla1}. 
The warm-starting variants use approximate solutions obtained from the Goemans-Williamson algorithm for
Max-Cut~\cite{gw-rounding}, greedy list scheduling for Partition~\cite{list-scheduling1} and the classic two-appro\-xi\-ma\-tion algorithm vor Vertex Cover~\cite{Cormen:2022}.

To compare algorithm performance, we consider \emph{approximation quality}, here defined as the ratio between the obtained solution
and the optimal solution for a problem-specific 
measure~(Max-Cut: size of the cut; 
Partition: cardinality of the smaller set;
Vertex Cover: reciprocal number of vertices in the cover\footnote{Invalid Vertex Cover solutions are replaced with the trivial vertex cover using the entire vertex set.}). A approximation quality of 1 corresponds to the optimal, and 0 to the worst possible solution.
Using density-matrix-based simulations is computationally more expensive than other alternatives, but allows us to obtain \emph{exact} values for $\bra\psi H_C \ket\psi$---the expected energy of the problem Hamiltonian---in noisy simulations.
For QAOA, WSQAOA and WS-Init-QAOA, we obtain the average value~(and average approximation quality) of the solution from the measurement probabilities of the  output state.
For RQAOA, this approach is not possible since the algorithm delivers only a single outcome per design, which we need to appropriately account for. 
During RQAOA, when using a reasonably small sample size, we cannot expect to always find the term $t$ with  maximum $|E[t]|$.
Therefore, we do not use the exact measurement probabilities to find the most \enquote{conclusive} term.
Instead, we consider a constant sample size of ten.
As shown below, this still achieves good results. 

\subsection{Noise Model}
The \href{https://qiskit.org/ecosystem/aer/_modules/qiskit_aer/noise/device/models.html}{Qiskit noise model} is widely 
used~\cite{noise-model,noise-model2,felix}, and
accounts for dominant noise in IBM quantum systems, featuring good agreement with actual quantum hardware~\cite{noise-model, noise-model-evaluation}.
We implemented this model on top of QLM's noisy circuit simulator,
but additionally allow for changing the strength of the individual noise sources.
Quantum channels describe
\begin{itemize}
    \item
    \emph{Gate errors:}
    Quantum gates can never be implemented perfectly.
    This manifests itself in the fact that instead of the desired unitary, a different, unknown unitary or non-unitary operator is applied instead.
    Gate errors are modelled using depolarising channels, which with some depolarising probability $p$ replace the state of the affected qubits by the maximally mixed state \cite{nielsen-chuang}.
    \item
    \emph{Thermal relaxation:}
    Even if the qubit is not involved in any quantum gates, its state slowly transitions to the thermal equilibrium state $\ket 0$ over time \cite{nielsen-chuang,superconducting-qubits}.
\end{itemize}

The model is parameterised by
\begin{enumerate*}[label=(\alph*)]
    \item longitudinal (\(T_{1}\)) and transverse (\(T_{2}\)) relaxation time,
    \item gate error and
    duration for each gate type (\eg $\rz$ or $\cnot$),
    \item strength of gate errors (\(d_{D}\)) and thermal relaxation (\(d_{\text{TR}}\)) as discussed in~\secref{sec:noise-levels}.
\end{enumerate*}
As a simplifying assumption, we assume all qubits share identical
imperfection characteristics.

\begin{table}[htbp]
    \centering
    \begin{tabular}{lrrrrr}
        \toprule
        & $\rz$ & $\sqrt X$ & $\cnot$ & \(T_{1}\) & \(T_{2}\)\\
        \midrule
        Gate error & \SI{0}{\percent} & \SI{0.03}{\percent} & \SI{1}{\percent} \\
        Duration/Time  & \SI{0}{\nano s} & \SI{35}{\nano s} & \SI{400}{\nano s} & \SI{100}{\micro s} & \SI{150}{\micro s} \\
        \bottomrule
    \end{tabular}
\caption{Baseline noise parameters.}\label{tab:noise-parameters}
\end{table}

\secref{tab:noise-parameters} shows our baseline parameters: They represent median values obtained from the 46 IBM QPUs whose data are available from \href{https://docs.quantum-computing.ibm.com/api/qiskit/providers_fake_provider}{fake backends}.
For each backend, we take the median over all qubits, and then obtain the median over backend medians to ensure that systems with more qubits are not disproportionately represented.
$T_1$ and $T_2$ are rounded to \SI{5}{\micro\second}, gate times to \SI{5}{\nano\second} and gate errors to one significant digit.
These values represent the current state of the art in transmonic NISQ devices. For lack of space, we cannot include other architectures, but our simulations can be easily performed for them using the \repro{}~\cite{mauerer-reproduction-package} (link in PDF).

The \emph{average fidelity} of a quantum channel measures how much it alters its input.
It lies in the interval $[2^{-n}, 1]$ where $n$ is the number of qubits the channel acts upon.
An average fidelity of $1$ means that the channel leaves its input unchanged, an average fidelity of $2^{-n}$ represents a complete loss of information.
Our model interprets the empirically determined gate error as one minus its average fidelity.
We assume this infidelity to be caused by both error types, which we model as depolarising noise and thermal relaxation.

After each logical single-qubit gate, we insert a thermal relaxation channel, parameterised by $T_1$, $T_2$ and gate duration $t$, followed by a depolarising channel.
The depolarising probability $p$ is set such that the average fidelity of the composed channel matches the model parameters
(the required probability can be derived analytically~\cite{average-fidelity,felix}).
After each two-qubit gate, one thermal relaxation channel is inserted for each involved qubit, followed by a two-qubit depolarising channel (again, we choose parameters so that the combined channel matches the desired fidelity).
Due to two-qubit gate dependencies, some qubits might not be subject to any gate during \emph{idle periods}. These are represented by a thermal relaxation channel parameterised by $T_1$, $T_2$, and idle duration.

We assume a native gate set $\big\{\rz, \sqrt{X}, \cnot\big\}$, as it is supported by many transmon devices, and transpile from logical to native circuits by substituting single-qubit gates.
While not all NISQ machines support $\cnot$ gates natively~(\eg the IBM Eagle), equivalents (up to single-qubit rotations) exist on all architectures. 
Since two-qubit gate imperfections usually exceeds single-qubit gate imperfections, any additional single-qubit gates that arise from substituting $\cnot$ should not meaningfully affect our results. $\rz$ gates are noiseless, as they are virtual on transmon systems~\cite{virtual-rz},
and Misra-Gries edge colouring~\cite{misra-gries} orders terms in \(H_{\text{C}}\) to minimise circuit depth.

\subsection{Strength of Noise}
\label{sec:noise-levels}
Recall that noise strengths are controlled via parameters $d_D$ and $d_{TR}$.
Specifically, $d_D$ scales the probability $p$ of all depolarising channels, $d_{TR}$ scales the time parameter $t$ of all thermal relaxation channels.
A value of 1 corresponds to the baseline noise level, whereas a value of 0 disables the noise source.
Intuitively, depolarising and thermal relaxation noise lead to a decrease of the state's quality (as measured by fidelity with the desired optimal state), which is exponential in the circuit's depth.
The average fidelity of the depolarising channel is $\nicefrac{1}{2} + \nicefrac{1}{2}\cdot e^{-t/T_1}$ if $T_1 = T_2$, 
and $d_D$ is a scaling factor of the decay constant ($1/T_1$)~\cite{nielsen-chuang,average-fidelity}.
Likewise, the average fidelity after $k$-fold application of the single-qubit depolarising channel with depolarisation probability $p$ is $\nicefrac{1}{2} + \nicefrac{1}{2}\cdot e^{k\ln(1 - p)}$~\cite{depolarizing-fidelity}.
As $\ln(1 - p) \approx p$ for $p \ll 1$, $d_D$ again scales the decay constant for small $p$.

\subsection{Runtime estimation}
To estimate circuit execution time, we use gate durations given in \autoref{tab:noise-parameters}, and a median measurement time of \SI{4.09}{\micro s} obtained from fake backends.
We assume 1000 circuit evaluations per iteration.
Measurements on the Qaptiva 800 determine efforts for parameter optimisation, circuit transpilation, or finding initial values for warm-starting variants.

\begin{figure}[htbp]
    \input{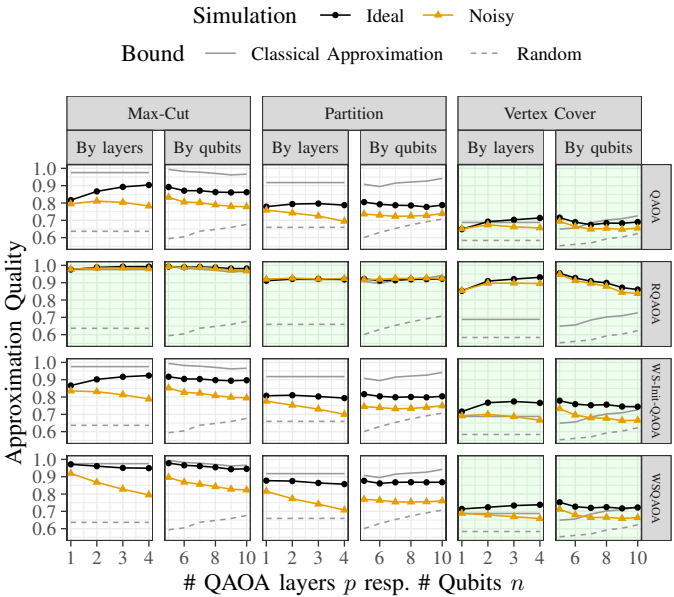}
    \vspace{-0.5em}
    \caption{
        Average approximation quality, separated by number of QAOA layers and averaged over number of qubits and vice versa, on random problem instances, for ideal and noisy evaluation.
        Dashed grey lines: lower bound (random guess). Solid grey lines: initial warm-start estimate.
        Green background: QAOA variant outperforms classical approximation algorithm.
    }
    \label{fig:performance}
\end{figure}
\section{Evaluation}\label{sec:evaluation}

\begin{figure*}
    \centering
    \input{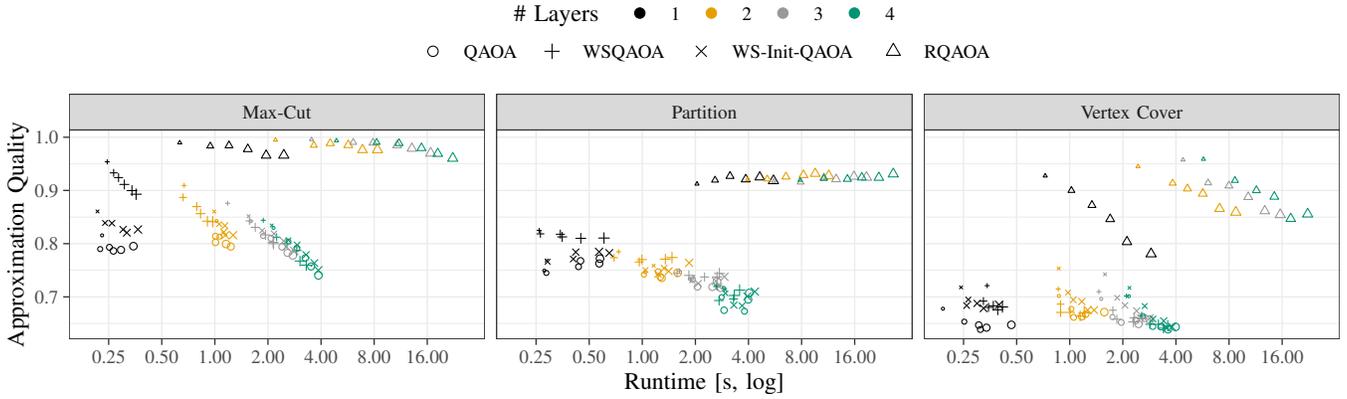}
    \caption{
    Runtime versus approximation quality for noisy simulation. Each point represents an average over 100 instances for a given number of qubits; point size increases with instance size respectively qubit count (five to ten).}
    \label{fig:performance_by_runtime}
\end{figure*}
Our core concern is to understand how non-functional properties of QAOA variants affect their integration into HPC systems, with a particular focus on the effect of imperfections on solution quality and runtime.
\secref{fig:performance} summarises our general observations: It 
shows average approximation quality, dependent on an increasing number of layers $p$ and averaged over the problem size $n$, as well as depending on $n$ and averaged over $p$, for two corner cases:
Full IBM Q device noise, and the noiseless case.
Overall, RQAOA performs best and standard QAOA worst in the ideal and the noisy cases. WSQAOA and WS-Init-QAOA ranked second and third, respectively. Although, as expected, all variants perform worse in the noisy case, there is a clear difference in performance deterioration;
particularly, RQAOA is more robust than other variants.
For an increasing number of layers, the better result quality in the noiseless case competes with a larger amount of sustained noise.
For IBM Q devices, detrimental effects usually exceed improvements, and using two QAOA layers instead of one shows only an improvement for standard QAOA and RQAOA. For the two warm-starting variants, the added power of a second layer does not outweigh the additional noise. While WSQAOA is slightly more affected, the effect of noise is mostly comparable across the other variants.
It appears reasonable that RQAOA is less disturbed by noise:
Instead of trying to approximate an optimal solution directly, RQAOA iteratively finds the most \enquote{conclusive} constraint and adds it to the Hamiltonian. This way, it finds the most highly 
correlated pair of variables in every step, which are then fixed
and not affected by future influence of noise.
WSQAOA struggles to benefit from adding more layers, even in the ideal case, which we attribute to the limitations of the relatively \enquote{weak} classical optimiser.

\autoref{fig:performance_by_runtime} compares solution quality and runtime for IBM Q noise levels. 
Each data point represents average approximation quality and runtime of an algorithm with $p \in [1, 4]$ layers for 100 problem instances with the same number of qubits. We observe that
\begin{enumerate*}[label=(\alph*)]
\item RQAOA is consistently best in terms of solution quality, albeit at the expense of runtime, since it executes QAOA multiple times,
\item differences in intra-algorithm execution times are mostly negligible for non-RQAOA variants, 
\item approximation quality decreases, for increasing problem dimension,
at markedly different rates depending on the subject problem, and
specific to each algorithm,
\item said decrease follows, from visual observation, a characteristic trend, 
\item distinct clusters arise that are connected to the amount of layers.
\end{enumerate*}
Assuming this behaviour remains valid for larger instances (which is, of course, not guaranteed~\cite{McGeoch2023:}), the clustering  presents an opportunity to select suitable execution parameters prior to payload execution.
\label{sec:noise-levels-evaluation}
\begin{figure}[htbp]
    \centering
    \input{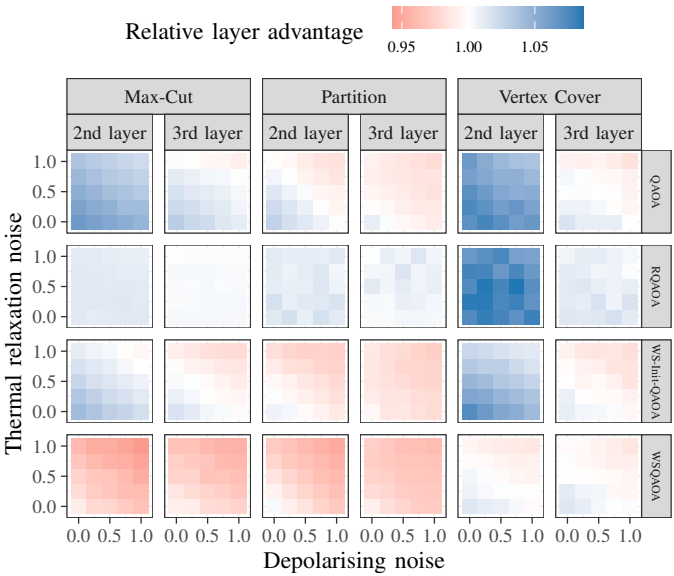}
    \tikzspace
    \caption{
        Relative advantage of a second and third QAOA layer depending on noise strength.
        Blue cells: performance increase; red cells: worse performance with added layer(s).
    }
    \label{fig:noise-levels}
\end{figure}
Finally, we consider the continuum between the noise-free case and IBM Q imperfection levels: \secref{fig:noise-levels} shows under
what conditions adding an extra layer improves performance.
We use relative advantage (\ie the ratio of $p$ layer performance over $p - 1$ layer performance) averaged over all subject problem instances.
For moderate noise, standard QAOA  benefits from adding a second or third layer, as does WS-Init-QAOA when adding a second. However, this
depends on the subject problem:
Partition is less resistant to noise than Max-Cut and Vertex Cover.
A likely reason is that circuits for Partition contain more $\cnot$ gates than those for Max-Cut or Vertex Cover instances:
For Partition, there is one $\cnot$ gate for every pair of qubits; for Max-Cut and Vertex Cover there is one $\cnot$ for each graph edge, which amounts to roughly half the pairs, since for each pair, an edge is inserted with \SI{50}{\percent} probability.

It is notable that in the majority of cases, relative advantage is consistently \emph{either} negative \emph{or} positive for a given combination of algorithm and problem. 
A change in trend depending on noise strength is rare.
This suggests that choosing the optimal number of layers does not need to take into account hardware noise details.

\section{Discussion \& Conclusion}
\label{sec:conclusion}
Our results show pronounced differences between the QAOA variants for solution quality, runtime or noise strength that can be pinpointed to
specific causes. This suggests that augmenting software stacks with capabilities to automatically select optimal QAOA variants with an appropriate number of layers for a given problem, based on non-functional criteria such as solution quality, latency, or quantum runtime is a possible and useful endeavour.
Our data suggest that in noisy environments, RQAOA has the potential to achieve better results than other variants, at the cost of a much longer runtime.
They also indicate that for most QAOA variants, the computational benefit of a second layer is often greatly reduced or even eliminated by the added noise of the deeper quantum circuits.
The considerable computational cost of simulating noisy quantum circuits allowed us to only consider problems with up to ten qubits.
Nonetheless, we believe that results for relatively small instances may indicate general trends, although this needs further confirmation.
If our expectation hold true, another possible approach to \mbox{(semi-)automatic} algorithm selection could be to probe smaller dummy instances at compile time or runtime to inform the selection for real problem instances.

Automatic algorithm selection needs to distinguish between problem-independent and problem-specific elements, and abstractions for deploying quantum algorithms such as QAOA in HPC systems need to take this into account.
While the trade-off between solution quality and runtime remains relatively consistent across the three subject problems when considering the algorithm and noise dimensions, the situation becomes more nuanced when examining different numbers of QAOA layers.
The point at which an additional layer does not yield performance improvements depends on the problem itself and on other factors like noise level or problem size.
Other aspects
---for instance, the impact of problem encodings \cite{Krueger:2020,Zielinski:2024} or noise caused by swap gates required because of limited connectivity \cite{safi:23:codesign}---, should also be considered.
More research beyond our results is needed to understand the effect of problem-specific aspects such as classical approximability on the optimal hyper-parameters of different QAOA variants.

To enjoy QC in HPC environments without the need for quantum training, the use of QAOA (and possibly 
other classes of algorithms) could be enabled by allowing users to specify relevant non-functional requirements (\eg acceptable solution quality or communication latency) in quantum code~\cite{schoenberger:22:icsa,Mauerer:2005,Lee:2023}, for instance via meta-annotations. If augmented with
appropriate (and already available~\cite{Lobe:2023}) means of providing objective functions
decoupled from quantum specifics, yet convertible to QUBO, and together with apt classical means of providing initial solution estimates, this would allow a compiler or runtime system to select the most suitable algorithm.
Our results suggest that this vision is achievable in the near future.

\newcommand{\WM}{\censor{WM}\xspace}
\newcommand{\ST}{\censor{ST}\xspace}
\newcommand{\HS}{\censor{HS}\xspace}
\newenvironment{acks}
  {\par\footnotesize\textbf{Acknowledgements:}}
  {\par\addvspace{\bigskipamount}}
\vspace*{0.25em}\begin{acks}
This work was supported by the German Federal Ministry of Education and 
Research (BMBF), within the funding program \enquote{quantum technologies—--from basic research 
to market}, grant numbers 13NI6092 (\WM, \ST) and 13N16093 (\HS). \WM and \ST acknowledge support by the
High-Tech Agenda of the Free State of Bavaria.
\end{acks}
\printbibliography

@article{error-correction,
author = {Joschka Roffe},
title = {Quantum error correction: an introductory guide},
journal = {Contemporary Physics},
volume = {60},
number = {3},
pages = {226-245},
year = {2019},
publisher = {Taylor & Francis},
doi = {10.1080/00107514.2019.1667078},
}

@article{nisq,
  doi       = {10.22331/q-2018-08-06-79},
  url       = {https://doi.org/10.22331%2Fq-2018-08-06-79},
  year      = 2018,
  month     = {aug},
  publisher = {Verein zur Forderung des Open Access Publizierens in den Quantenwissenschaften},
  volume    = {2},
  pages     = {79},
  author    = {John Preskill},
  title     = {Quantum Computing in the {NISQ} era and beyond},
  journal   = {Quantum}
}

@misc{qaoa,
  title         = {A Quantum Approximate Optimization Algorithm},
  author        = {Edward Farhi and Jeffrey Goldstone and Sam Gutmann},
  year          = {2014},
  eprint        = {1411.4028},
  archiveprefix = {arXiv},
  primaryclass  = {quant-ph}
}

@article{wsqaoa,
  doi       = {10.22331/q-2021-06-17-479},
  url       = {https://doi.org/10.22331\%2Fq-2021-06-17-479},
  year      = 2021,
  month     = {jun},
  publisher = {Verein zur Forderung des Open Access Publizierens in den Quantenwissenschaften},
  volume    = {5},
  pages     = {479},
  author    = {Daniel J. Egger and Jakub Mare{\v{c}}ek and Stefan Woerner},
  title     = {Warm-starting quantum optimization},
  journal   = {Quantum}
}

@article{gw-rounding,
  author     = {Goemans, Michel X. and Williamson, David P.},
  title      = {Improved Approximation Algorithms for Maximum Cut and Satisfiability Problems Using Semidefinite Programming},
  year       = {1995},
  issue_date = {Nov. 1995},
  publisher  = {Association for Computing Machinery},
  volume     = {42},
  number     = {6},
  issn       = {0004-5411},
  url        = {https://doi.org/10.1145/227683.227684},
  doi        = {10.1145/227683.227684},
  journal    = {J. ACM},
  month      = {nov},
  pages      = {1115–1145},
  numpages   = {31}
}

@article{rqaoa,
author = {Bravyi, Sergey and Kliesch, Alexander and Koenig, Robert and Tang, Eugene},
year = {2020},
month = {12},
pages = {},
title = {Obstacles to Variational Quantum Optimization from Symmetry Protection},
volume = {125},
journal   = {Phys. Rev. Lett.},
doi = {10.1103/PhysRevLett.125.260505}
}

@book{nielsen-chuang,
  place     = {Cambridge},
  title     = {Quantum Computation and Quantum Information},
  doi       = {10.1017/CBO9780511976667},
  publisher = {Cambridge University Press},
  author    = {Nielsen, Michael A. and Chuang, Isaac L.},
  year      = {2010}
}

@article{depolarizing-fidelity,
  author     = {Magesan, Easwar},
  title      = {Depolarizing Behavior of Quantum Channels in Higher Dimensions},
  year       = {2011},
  issue_date = {May 2011},
  publisher  = {Rinton Press, Incorporated},
  doi = {https://dl.acm.org/doi/10.5555/2011406.2011414},  
  volume     = {11},
  number     = {5},
  issn       = {1533-7146},
  journal    = {Quantum Info. Comput.},
  month      = {may},
  x-pages      = {466–484},
  numpages   = {19}
}

@article{ising,
  title   = {Ising formulations of many {NP} problems},
  volume  = {2},
  issn    = {2296-424X},
  url     = {http://journal.frontiersin.org/article/10.3389/fphy.2014.00005/abstract},
  doi     = {10.3389/fphy.2014.00005},
  urldate = {2023-08-25},
  journal = {Frontiers in Physics},
  author  = {Lucas, Andrew},
  year    = {2014}
}

@InProceedings{wsinitqaoa,
author="Awasthi, Abhishek
and B{\"a}r, Francesco
and Doetsch, Joseph
and Ehm, Hans
and Erdmann, Marvin
and Hess, Maximilian
and Klepsch, Johannes
and Limacher, Peter A.
and Luckow, Andre
and Niedermeier, Christoph
and Palackal, Lilly
and Pfeiffer, Ruben
and Ross, Philipp
and Safi, Hila
and Sch{\"o}nmeier-Kromer, Janik
and von Sicard, Oliver
and Wenger, Yannick
and Wintersperger, Karen
and Yarkoni, Sheir",
x-editor="Arai, Kohei",
title="Quantum Computing Techniques for Multi-knapsack Problems",
doi="https://doi.org/10.1007/978-3-031-37963-5_19",
booktitle="Intelligent Computing",
year="2023",
publisher="Springer Nature",
x-pages="264--284",
isbn="978-3-031-37963-5"
}

@article{rqaoa2,
	title = {Recursive {QAOA} outperforms the original {QAOA} for the {MAX}-{CUT} problem on complete graphs},
	volume = {23},
	issn = {1573-1332},
	url = {https://link.springer.com/10.1007/s11128-024-04286-0},
	doi = {10.1007/s11128-024-04286-0},
	number = {3},
	urldate = {2024-07-17},
	journal = {Quantum Information Processing},
	author = {Bae, Eunok and Lee, Soojoon},
	month = feb,
	year = {2024},
	pages = {78},
}

@article{average-fidelity,
  doi       = {10.1016/s0375-9601(02)01272-0},
  url       = {https://doi.org/10.1016%2Fs0375-9601%2802%2901272-0},
  year      = 2002,
  month     = {oct},
  publisher = {Elsevier {BV}},
  volume    = {303},
  x-number    = {4},
  x-pages     = {249--252},
  author    = {Michael A Nielsen},
  title     = {A simple formula for the average gate fidelity of a quantum dynamical operation},
  journal   = {Physics Letters A}
}

@inproceedings{felix,
 author = {Greiwe, Felix and Krüger, Tom and Mauerer, Wolfgang},
 booktitle = {Proc.\ IEEE International Conference on Quantum Software},
 doi = {10.1109/QSW59989.2023.00014},
 entrysubtype = {Conference},
 x-pages = {31-42},
 title = {Effects of Imperfections on Quantum Algorithms: A Software Engineering Perspective},
 url = {https://doi.org/10.1109/QSW59989.2023.00014},
 userd = {IEEE QSW '23},
 year = {2023}
}

@article{list-scheduling1,
  author={Graham, R. L.},
  journal={The Bell System Technical Journal}, 
  title={Bounds for certain multiprocessing anomalies}, 
  year={1966},
  volume={45},
  number={9},
  pages={1563-1581},
  doi={10.1002/j.1538-7305.1966.tb01709.x}
}

@article{superconducting-qubits,
  doi       = {10.1063/1.5089550},
  url       = {https://doi.org/10.1063%2F1.5089550},
  year      = 2019,
  month     = {jun},
  publisher = {{AIP} Publishing},
  volume    = {6},
  number    = {2},
  author    = {P. Krantz and M. Kjaergaard and F. Yan and T. P. Orlando and S. Gustavsson and W. D. Oliver},
  title     = {A quantum engineer{\textquotesingle}s guide to superconducting qubits},
  journal   = {Applied Physics Reviews}
}

@article{noise-model,
  title   = {Modeling and simulating the noisy behavior of near-term quantum computers},
  issn    = {2469-9926, 2469-9934},
  url     = {https://link.aps.org/doi/10.1103/PhysRevA.104.062432},
  doi     = {10.1103/PhysRevA.104.062432},
  urldate = {2023-10-01},
  journal = {PRA},
  author  = {Georgopoulos, Konstantinos and Emary, Clive and Zuliani, Paolo},
  month   = dec,
  year    = {2021}
}

@article{noise-model2,
  title    = {Quantum classifier with tailored quantum kernel},
  volume   = {6},
  issn     = {2056-6387},
  url      = {https://www.nature.com/articles/s41534-020-0272-6},
  doi      = {10.1038/s41534-020-0272-6},
  number   = {1},
  urldate  = {2023-10-01},
  journal  = {npj Quantum Information},
  author   = {Blank, Carsten and Park, Daniel K. and Rhee, June-Koo Kevin and Petruccione, Francesco},
  month    = may,
  year     = {2020},
  pages    = {41}
}

@article{virtual-rz,
  doi       = {10.1103/physreva.96.022330},
  url       = {https://doi.org/10.1103%2Fphysreva.96.022330},
  year      = 2017,
  month     = {aug},
  publisher = {American Physical Society ({APS})},
  volume    = {96},
  number    = {2},
  author    = {David C. McKay and Christopher J. Wood and Sarah Sheldon and Jerry M. Chow and Jay M. Gambetta},
  title     = {{Efficient Z-Gates for Quantum Computing}},
  journal   = {Physical Review A}
}

@incollection{cobyla1,
  title     = {A {Direct} {Search} {Optimization} {Method} {That} {Models} the {Objective} and {Constraint} {Functions} by {Linear} {Interpolation}},
  isbn      = {9789048143580},
  url       = {http://link.springer.com/10.1007/978-94-015-8330-5_4},
  booktitle = {Advances in {Optimization} and {Numerical} {Analysis}},
  publisher = {Springer},
  author    = {Powell, M. J. D.},
  x-editor    = {Gomez, Susana and Hennart, Jean-Pierre},
  year      = {1994},
  doi       = {10.1007/978-94-015-8330-5_4},
  x-pages     = {51--67}
}

@article{noisy-qaoa,
doi = {10.1088/0256-307X/38/3/030302},
url = {https://dx.doi.org/10.1088/0256-307X/38/3/030302},
year = {2021},
month = {mar},
publisher = {Chinese Physical Society and IOP Publishing Ltd},
volume = {38},
author = {Cheng Xue and Zhao-Yun Chen and Yu-Chun Wu and Guo-Ping Guo},
title = {Effects of Quantum Noise on Quantum Approximate Optimization Algorithm},
journal = {Chinese Physics Letters}
}

@article{noisy-qaoa2,
  doi       = {10.1088/2633-1357/abb0d7},
  url       = {https://doi.org/10.1088%2F2633-1357%2Fabb0d7},
  year      = 2020,
  month     = {aug},
  publisher = {{IOP} Publishing},
  volume    = {1},
  number    = {2},
  x-pages     = {025208},
  author    = {Jeffrey Marshall and Filip Wudarski and Stuart Hadfield and Tad Hogg},
  title     = {Characterizing local noise in {QAOA} circuits},
  journal   = {{IOP} {SciNotes}}
}

@article{scipy,
  author  = {Virtanen, Pauli and Gommers, Ralf and Oliphant, Travis E. and
             Haberland, Matt and Reddy, Tyler and Cournapeau, David and
             Burovski, Evgeni and Peterson, Pearu and Weckesser, Warren and
             Bright, Jonathan and {van der Walt}, St{\'e}fan J. and
             Brett, Matthew and Wilson, Joshua and Millman, K. Jarrod and
             Mayorov, Nikolay and Nelson, Andrew R. J. and Jones, Eric and
             Kern, Robert and Larson, Eric and Carey, C J and
             Polat, {\.I}lhan and Feng, Yu and Moore, Eric W. and
             {VanderPlas}, Jake and Laxalde, Denis and Perktold, Josef and
             Cimrman, Robert and Henriksen, Ian and Quintero, E. A. and
             Harris, Charles R. and Archibald, Anne M. and
             Ribeiro, Ant{\^o}nio H. and Pedregosa, Fabian and
             {van Mulbregt}, Paul and {SciPy 1.0 Contributors}},
  title   = {{{SciPy} 1.0: Fundamental Algorithms for Scientific
             Computing in Python}},
  journal = {Nature Methods},
  year    = {2020},
  volume  = {17},
  pages   = {261--272},
  adsurl  = {https://rdcu.be/b08Wh},
  doi     = {10.1038/s41592-019-0686-2}
}

@article{mauerer-reproduction-package,
  title={1-2-3 Reproducibility for Quantum Software Experiments},
  author={Wolfgang Mauerer and Stefanie Scherzinger},
  journal={IEEE SANER},
  doi = {10.1109/SANER53432.2022.00148},
  year={2022},
  pages={1247-1248}
}

@article{noise-model-evaluation,
doi = {10.1088/1402-4896/ad406c},
url = {https://dx.doi.org/10.1088/1402-4896/ad406c},
year = {2024},
month = {may},
publisher = {IOP Publishing},
volume = {99},
number = {6},
author = {Tom Weber and Kerstin Borras and Karl Jansen and Dirk Krücker and Matthias Riebisch},
title = {Construction and volumetric benchmarking of quantum computing noise models},
journal = {Physica Scripta}
}

@article{vertex-cover-approximation,
author = {Karakostas, George},
title = {A better approximation ratio for the vertex cover problem},
year = {2009},
issue_date = {October 2009},
publisher = {Association for Computing Machinery},
address = {New York, NY, USA},
volume = {5},
number = {4},
issn = {1549-6325},
url = {https://doi.org/10.1145/1597036.1597045},
doi = {10.1145/1597036.1597045},
journal = {ACM Trans. Algorithms},
month = {nov},
articleno = {41},
numpages = {8}
}

@article{misra-gries,
title = {A constructive proof of Vizing's theorem},
journal = {Information Processing Letters},
volume = {41},
number = {3},
pages = {131-133},
year = {1992},
issn = {0020-0190},
doi = {https://doi.org/10.1016/0020-0190(92)90041-S},
url = {https://www.sciencedirect.com/science/article/pii/002001909290041S},
author = {J. Misra and David Gries}
}

@article{McGeoch2023:,
author = {McGeoch, Catherine C. and Farr\'{e}, Pau},
title = {Milestones on the Quantum Utility Highway: Quantum Annealing Case Study},
year = {2023},
issue_date = {March 2024},
publisher = {Association for Computing Machinery},
address = {New York, NY, USA},
volume = {5},
url = {https://doi.org/10.1145/3625307},
doi = {10.1145/3625307},
journal = {ACM TQC},
month = {dec},
articleno = {2},
numpages = {30}
}

@inproceedings{Hoefler:2015,
author = {Hoefler, Torsten and Belli, Roberto},
title = {Scientific benchmarking of parallel computing systems: twelve ways to tell the masses when reporting performance results},
year = {2015},
isbn = {9781450337236},
publisher = {ACM},
url = {https://doi.org/10.1145/2807591.2807644},
doi = {10.1145/2807591.2807644},
booktitle = {Proc.\ Int.\ Conf.\ for HPC, Networking, Storage and Analysis},
articleno = {73},
numpages = {12},
}

@article{subset-sum-fptas,
author = {Ibarra, Oscar H. and Kim, Chul E.},
title = {Fast Approximation Algorithms for the Knapsack and Sum of Subset Problems},
year = {1975},
issue_date = {Oct. 1975},
publisher = {Association for Computing Machinery},
address = {New York, NY, USA},
volume = {22},
number = {4},
issn = {0004-5411},
url = {https://doi.org/10.1145/321906.321909},
doi = {10.1145/321906.321909},
journal = {J. ACM},
month = {oct},
x-pages = {463–468},
numpages = {6}
}

@article{Wang:2020,
  title = {$XY$ mixers: Analytical and numerical results for the quantum alternating operator ansatz},
  author = {Wang, Zhihui and Rubin, Nicholas C. and Dominy, Jason M. and Rieffel, Eleanor G.},
  journal = {Phys. Rev. A},
  volume = {101},
  issue = {1},
  numpages = {16},
  year = {2020},
  month = {Jan},
  publisher = {American Physical Society},
  doi = {10.1103/PhysRevA.101.012320},
  url = {https://link.aps.org/doi/10.1103/PhysRevA.101.012320}
}

@article{Zhu:2022,
  title = {Adaptive quantum approximate optimization algorithm for solving combinatorial problems on a quantum computer},
  author = {Zhu, Linghua and Tang, Ho Lun and Barron, George S. and Calderon-Vargas, F. A. and Mayhall, Nicholas J. and Barnes, Edwin and Economou, Sophia E.},
  journal = {Phys. Rev. Res.},
  volume = {4},
  numpages = {9},
  year = {2022},
  month = {Jul},
  publisher = {American Physical Society},
  doi = {10.1103/PhysRevResearch.4.033029},
  url = {https://link.aps.org/doi/10.1103/PhysRevResearch.4.033029}
}

@INPROCEEDINGS{Baertschi:2020,
  author={Bärtschi, Andreas and Eidenbenz, Stephan},
  booktitle={Proc.\ IEEE QCE}, 
  title={Grover {Mixers} for {QAOA}: Shifting Complexity from Mixer Design to State Preparation}, 
  year={2020},
  volume={},
  number={},
  pages={72-82},
  doi={10.1109/QCE49297.2020.00020}
}

@article{Zhang:2017,
  title = {Near-optimal quantum circuit for Grover's unstructured search using a transverse field},
  author = {Jiang, Zhang and Rieffel, Eleanor G. and Wang, Zhihui},
  journal = {Phys.\ Rev.\ A},
  volume = {95},
  issue = {6},
  numpages = {8},
  year = {2017},
  month = {Jun},
  publisher = {American Physical Society},
  doi = {10.1103/PhysRevA.95.062317},
  url = {https://link.aps.org/doi/10.1103/PhysRevA.95.062317}
}

@article{Tate:2023,
  doi = {10.22331/q-2023-09-26-1121},
  url = {https://doi.org/10.22331/q-2023-09-26-1121},
  title = {Warm-{S}tarted {QAOA} with {C}ustom {M}ixers {P}rovably {C}onverges and {C}omputationally {B}eats {G}oemans-{W}illiamson's {M}ax-{C}ut at {L}ow {C}ircuit {D}epths},
  author = {Tate, Reuben and Moondra, Jai and Gard, Bryan and Mohler, Greg and Gupta, Swati},
  journal = {{Quantum}},
  issn = {2521-327X},
  publisher = {{Verein zur F{\"{o}}rderung des Open Access Publizierens in den Quantenwissenschaften}},
  volume = {7},
  pages = {1121},
  month = sep,
  year = {2023}
}

@article{Sud:2024,
  title = {Parameter-setting heuristic for the quantum alternating operator ansatz},
  author = {Sud, James and Hadfield, Stuart and Rieffel, Eleanor and Tubman, Norm and Hogg, Tad},
  journal = {Phys. Rev. Res.},
  volume = {6},
  year = {2024},
  month = {May},
  publisher = {American Physical Society},
  doi = {10.1103/PhysRevResearch.6.023171},
  url = {https://link.aps.org/doi/10.1103/PhysRevResearch.6.023171}
}

@misc{montanezbarrera:2024,
      title={Transfer learning of optimal QAOA parameters in combinatorial optimization}, 
      author={J. A. Montanez-Barrera and Dennis Willsch and Kristel Michielsen},
      year={2024},
      eprint={2402.05549},
      url={https://arxiv.org/abs/2402.05549}, 
      doi={https://doi.org/10.48550/arXiv.2402.05549},
}

@article{Streif:2020,
	title = {Training the quantum approximate optimization algorithm without access to a quantum processing unit},
	volume = {5},
	issn = {2058-9565},
	url = {https://iopscience.iop.org/article/10.1088/2058-9565/ab8c2b},
	doi = {10.1088/2058-9565/ab8c2b},
	urldate = {2024-07-17},
	journal = {Quantum Science and Technology},
	author = {Streif, Michael and Leib, Martin},
	month = may,
	year = {2020},
}

@article{Vijendran:2024,
author = {Vijendran, V and Das, Aritra and Koh, Dax and Assad, Syed and Lam, Ping Koy},
year = {2024},
month = {02},
pages = {},
title = {An Expressive Ansatz for Low-Depth Quantum Approximate Optimisation},
volume = {9},
journal = {Quantum Sci.\ and Tech.},
doi = {10.1088/2058-9565/ad200a}
}

@article{Humble:2021,
  author={Humble, Travis S. and McCaskey, Alexander and Lyakh, Dmitry I. and Gowrishankar, Meenambika and Frisch, Albert and Monz, Thomas},
  journal={IEEE Micro}, 
  title={Quantum Computers for High-Performance Computing}, 
  year={2021},
  volume={41},
  number={5},
  pages={15-23},
  doi={10.1109/MM.2021.3099140}
}

@article{Karalekas:2020,
doi = {10.1088/2058-9565/ab7559},
url = {https://dx.doi.org/10.1088/2058-9565/ab7559},
year = {2020},
month = {mar},
publisher = {IOP Publishing},
volume = {5},
number = {2},
author = {Peter J Karalekas and Nikolas A Tezak and Eric C Peterson and Colm A Ryan and Marcus P da Silva and Robert S Smith},
title = {A quantum-classical cloud platform optimized for variational hybrid algorithms},
journal = {Q.\ Sci.\ \& Tech.},
}

@incollection{Wintersperger:2022,
	title = {{QPU}-{System} {Co}-design for {Quantum} {HPC} {Accelerators}},
	isbn = {9783031218668},
	url = {https://link.springer.com/10.1007/978-3-031-21867-5_7},
	urldate = {2024-07-17},
	booktitle = {Proc.\ ARCS},
	author = {Wintersperger, Karen and Safi, Hila and Mauerer, Wolfgang},
	year = {2022},
	doi = {10.1007/978-3-031-21867-5_7},
	pages = {100--114},
}

@book{Cormen:2022,
	address = {Cambridge, Massachusett},
	edition = {Fourth edition},
	title = {Introduction to algorithms},
	isbn = {9780262046305},
	publisher = {The MIT Press},
	author = {Cormen, Thomas H. and Leiserson, Charles Eric and Rivest, Ronald L. and Stein, Clifford},
	year = {2022},
}

@inproceedings{Lobe:2023,
author = "Lobe, Elisabeth",
title = "quark: QUantum Application Reformulation Kernel",
year = 2023,
doi = "10.18420/inf2023_123",
booktitle = "INFORMATIK 2023",
publisher = "GI e.V.",
address = "Bonn",
pissn = "1617-5468",
isbn = "978-3-88579-731-9",
pages = "1115--1120",
}

@misc{Elsharkawy:2023,
	title = {Integration of {Quantum} {Accelerators} with {High} {Performance} {Computing} -- {A} {Review} of {Quantum} {Programming} {Tools}},
	url = {http://arxiv.org/abs/2309.06167},
	doi = {10.48550/arXiv.2309.06167},
	urldate = {2024-07-17},
	publisher = {arXiv},
	author = {Elsharkawy, Amr and To, Xiao-Ting Michelle and Seitz, Philipp and Chen, Yanbin and Stade, Yannick and Geiger, Manuel and Huang, Qunsheng and Guo, Xiaorang and Ansari, Muhammad Arslan and Mendl, Christian B. and Kranzlmüller, Dieter and Schulz, Martin},
	month = sep,
	year = {2023},
	note = {arXiv:2309.06167 [quant-ph]},
}

@inproceedings{Bandic:2022,
	title = {Full-stack quantum computing systems in the {NISQ} era: algorithm-driven and hardware-aware compilation techniques},
	isbn = {9783981926361},
	shorttitle = {Full-stack quantum computing systems in the {NISQ} era},
	urldate = {2024-07-17},
	booktitle = {Proc.\ DAAD},
	author = {Bandic, Medina and Feld, Sebastian and Almudever, Carmen G.},
	month = may,
	year = {2022},
	pages = {1--6},
    doi = {10.23919/DATE54114.2022.9774643}
}

@article{Alexeev:2024,
title = {Quantum-centric supercomputing for materials science: A perspective on challenges and future directions},
journal = {Future Generation Computer Systems},
volume = {160},
pages = {666-710},
year = {2024},
issn = {0167-739X},
doi = {https://doi.org/10.1016/j.future.2024.04.060},
url = {https://www.sciencedirect.com/science/article/pii/S0167739X24002012},
author = {Yuri Alexeev and Maximilian Amsler and Marco Antonio Barroca and Sanzio Bassini and Torey Battelle and Daan Camps and David Casanova and Young Jay Choi and Frederic T. Chong and Charles Chung and Christopher Codella and Antonio D. Córcoles and James Cruise and Alberto {Di Meglio} and Ivan Duran and Thomas Eckl and Sophia Economou and Stephan Eidenbenz and Bruce Elmegreen and Clyde Fare and Ismael Faro and Cristina Sanz Fernández and Rodrigo Neumann Barros Ferreira and Keisuke Fuji and Bryce Fuller and Laura Gagliardi and Giulia Galli and Jennifer R. Glick and Isacco Gobbi and Pranav Gokhale and Salvador {de la Puente Gonzalez} and Johannes Greiner and Bill Gropp and Michele Grossi and Emanuel Gull and Burns Healy and Matthew R. Hermes and Benchen Huang and Travis S. Humble and Nobuyasu Ito and Artur F. Izmaylov and Ali Javadi-Abhari and Douglas Jennewein and Shantenu Jha and Liang Jiang and Barbara Jones and Wibe Albert {de Jong} and Petar Jurcevic and William Kirby and Stefan Kister and Masahiro Kitagawa and Joel Klassen and Katherine Klymko and Kwangwon Koh and Masaaki Kondo and Dog̃a Murat Kürkçüog̃lu and Krzysztof Kurowski and Teodoro Laino and Ryan Landfield and Matt Leininger and Vicente Leyton-Ortega and Ang Li and Meifeng Lin and Junyu Liu and Nicolas Lorente and Andre Luckow and Simon Martiel and Francisco Martin-Fernandez and Margaret Martonosi and Claire Marvinney and Arcesio Castaneda Medina and Dirk Merten and Antonio Mezzacapo and Kristel Michielsen and Abhishek Mitra and Tushar Mittal and Kyungsun Moon and Joel Moore and Sarah Mostame and Mario Motta and Young-Hye Na and Yunseong Nam and Prineha Narang and Yu-ya Ohnishi and Daniele Ottaviani and Matthew Otten and Scott Pakin and Vincent R. Pascuzzi and Edwin Pednault and Tomasz Piontek and Jed Pitera and Patrick Rall and Gokul Subramanian Ravi and Niall Robertson and Matteo A.C. Rossi and Piotr Rydlichowski and Hoon Ryu and Georgy Samsonidze and Mitsuhisa Sato and Nishant Saurabh and Vidushi Sharma and Kunal Sharma and Soyoung Shin and George Slessman and Mathias Steiner and Iskandar Sitdikov and In-Saeng Suh and Eric D. Switzer and Wei Tang and Joel Thompson and Synge Todo and Minh C. Tran and Dimitar Trenev and Christian Trott and Huan-Hsin Tseng and Norm M. Tubman and Esin Tureci and David García Valiñas and Sofia Vallecorsa and Christopher Wever and Konrad Wojciechowski and Xiaodi Wu and Shinjae Yoo and Nobuyuki Yoshioka and Victor Wen-zhe Yu and Seiji Yunoki and Sergiy Zhuk and Dmitry Zubarev},
}

@inproceedings{Krueger:2020,
author = {Kr\"{u}ger, Tom and Mauerer, Wolfgang},
title = {Quantum Annealing-Based Software Components: An Experimental Case Study with SAT Solving},
year = {2020},
isbn = {9781450379632},
publisher = {ACM},
address = {New York, NY, USA},
url = {https://doi.org/10.1145/3387940.3391472},
doi = {10.1145/3387940.3391472},
booktitle = {Proc.\ ICSEW},
pages = {445–450},
numpages = {6},
}

@inproceedings{safi:23:codesign,
 author = {Safi, Hila and Wintersperger, Karen and Mauerer, Wolfgang},
 booktitle = {IEEE International Conference on Quantum Software},
 doi = {10.1109/QSW59989.2023.00022},
 pages = {104-115},
 title = {Influence of HW-SW-Co-Design on Quantum Computing Scalability},
 year = {2023}
}

@INPROCEEDINGS{Farooqi:2023,
  author={Farooqi, Muhammad Nufail and Ruefenacht, Martin},
  booktitle={IEEE QCE}, 
  title={Exploring Hybrid Classical-Quantum Compute Systems through Simulation}, 
  year={2023},
  volume={02},
  number={},
  pages={127-133},
  doi={10.1109/QCE57702.2023.10196}
}

@inproceedings{Zielinski:2024,
title = "{SATQUBOLIB}: A Python Framework for Creating and Benchmarking {(Max-)3SAT QUBOs}",
author = "Sebastian Zielinski and Magdalena Benkard and Jonas N{\"u}{\ss}lein and Claudia Linnhoff-Popien and Sebastian Feld",
year = "2024",
doi = "10.1007/978-3-031-60433-1_4",
isbn = "978-3-031-60432-4",
publisher = "Springer",
pages = "48--66",
booktitle = "Proc.\ I4CS",
}

@article{Mauerer:2005,
  title={Semantics and simulation of communication in quantum programming},
  author={Mauerer, Wolfgang},
  journal={arXiv preprint quant-ph/0511145},
  year={2005},
  doi={https://doi.org/10.48550/arXiv.quant-ph/0511145},
}

@INPROCEEDINGS {Campbell:2023,
author = {C. Campbell and F. T. Chong and D. Dahl and P. Frederick and P. Goiporia and P. Gokhale and B. Hall and S. Issa and E. Jones and S. Lee and A. Litteken and V. Omole and D. Owusu-Antwi and M. A. Perlin and R. Rines and K. N. Smith and N. Goss and A. Hashim and R. Naik and E. Younis and D. Lobser and C. G. Yale and B. Huang and J. Liu},
booktitle = {IEEE QCE},
title = {Superstaq: Deep Optimization of Quantum Programs},
year = {2023},
volume = {},
issn = {},
pages = {1020-1032},
doi = {10.1109/QCE57702.2023.00116},
url = {https://doi.ieeecomputersociety.org/10.1109/QCE57702.2023.00116},
month = {sep}
}

@inproceedings{schoenberger:22:icsa,
 author = {Schönberger, Manuel and Franz, Maja and Scherzinger, Stefanie and Mauerer, Wolfgang},
 booktitle = {IEEE ICSA-C},
 doi = {10.1109/ICSA-C54293.2022.00039},
 entrysubtype = {Workshop},
 pages = {164-169},
 title = {Peel \(\mid\) Pile? Cross-Framework Portability of Quantum Software},
 url = {https://ieeexplore.ieee.org/document/9779833},
 userc = {CORE22:A},
 userd = {ICSA '22},
 year = {2022}
}

@INPROCEEDINGS{Lee:2023,
title = "Quantum Task Offloading with the OpenMP API",
author = "Joseph Lee and Oliver Brown and Bull,{Jonathan Mark} and Martin Ruefenacht and Johannes Doerfert and Michael Klemm and Martin Schulz",
year = "2023",
 booktitle = "SC23",
}
\end{document}